\documentclass{svproc}
\usepackage{url}
\usepackage{graphicx}

\begin{document}
\mainmatter              % start of a contribution
\title{Production of hypernuclei and properties of hyper-nuclear matter}
\titlerunning{Hamiltonian Mechanics}  % abbreviated title (for running head)
\author{Alexander Botvina\inst{1} \and  Marcus Bleicher\inst{2} \and 
Nihal Buyukcizmeci\inst{3}}
\authorrunning{Alexander Botvina et al.} % abbreviated author list (for running head)
\tocauthor{Alexander Botvina, Marcus Bleicher, Nihal Buyukcizmeci}
\institute{ITP and FIAS, J.W.Goethe University, Frankfurt am Main, Germany;\\
INR, Russian Academy of Sciences, Moscow, Russia.\\ 
\and
ITP and FIAS, J.W.Goethe University, Frankfurt am Main, Germany;\\
GSI Helmholtz Center for Heavy Ion Research, Darmstadt, Germany;\\
John-von-Neumann Institute for Computing (NIC), FZ Julich, Julich, Germany.\\
\and
Department of Physics, Selcuk University, 42079 Kampus, Konya, Turkey.}

\maketitle              % typeset the title of the contribution

\begin{abstract}
The relativistic nucleus-nucleus collisions can produce hypernuclei and 
low-temperature hyper-matter as a result of hyperon capture by 
nuclear residues and free nucleons. 
We use the transport, coalescence and statistical models to describe the 
whole process, and point at the important advantages of such reactions: 
A broad variety of formed hypernuclei in masses and 
isospin allows for investigating properties of exotic hypernuclei, 
as well as the hypermatter both at high and low temperatures. The 
abundant production of multi-strange nuclei that can give an access to 
multi-hyperon systems and strange nuclear matter. The de-excitation of 
hot hyper-cluster will allow for the hyperon correlation studies. 
There is a saturation 
of the hypernuclei production at high energies, therefore, the optimal way 
to pursue this experimental study is to use the accelerator facilities of 
intermediate energies. 
%\dots
% We would like to encourage you to list your keywords within
% the abstract section using the \keywords{...} command.
\end{abstract}
%
%\section{Studies of hypernuclei in relativistic ion collisions}

Embedding hyperons in 
the nuclear matter allows to explore the many-body aspects of the strong 
three-flavor interaction (i.e., including $u$, $d$, and $s$ quarks) at 
low energies. Heavy hypernuclei open also opportunities to study the 
hyperon interactions and properties of strange matter, 
that is important for nuclear astrophysics (e.g., in neutron stars). 
Many various hyperons ($\Lambda,\Sigma,\Xi,\Omega$) are produced in 
relativistic nuclear collisions. These hyperons can be captured by 
the produced baryons as well as by the projectile/target 
nuclear residues with the formation of hypernuclei.  
In such deep-inelastic reactions leading to fragmentation processes one 
can form hypernuclei of all sizes and isospin content, that gives 
advantages over traditional experimental hypernuclear methods. There are 
many experimental collaborations (STAR at RHIC, ALICE at LHC, PANDA, CBM, 
HypHI, Super-FRS, R3B at GSI/FAIR, BM@N, MPD at NICA) which plan to 
investigate hypernuclei and their properties in reactions induced by 
relativistic ions. The isospin space, particle unstable states, 
multi-strange nuclei and their precise lifetime can be explored in these 
fragmentation reactions. 
It was theoretically demonstrated with numerous models 
\cite{Cas95,Bot07,giessen,Bot11,Bot13,Buy13,Bot15,Bot17} 
that in nuclear collisions one can produce 
all kind of hypernuclei including multi-strange and exotic ones. 
Moreover the properties of the hypernuclei (e.g., the hyperon 
binding energies) can be extracted directly from their yields 
\cite{Buy18}. 
There also exist experimental confirmations of such processes leading to 
light hypernuclei \cite{saito-new,star,alice} and fission hypernuclei 
\cite{Arm93,Ohm97}. 

\section{Formation of hypernuclei in relativistic ion collisions}
%
%\subsection{Autonomous Systems}
%

% my figure 1
% Figure1
\vspace{-8mm}
%\begin{figure}
\begin{figure}[tbh]
%\hspace{-72mm}
\hspace{-5mm}
%\begin{center}
\begin{minipage}[t]{-72mm}
\includegraphics[width=7.2cm]{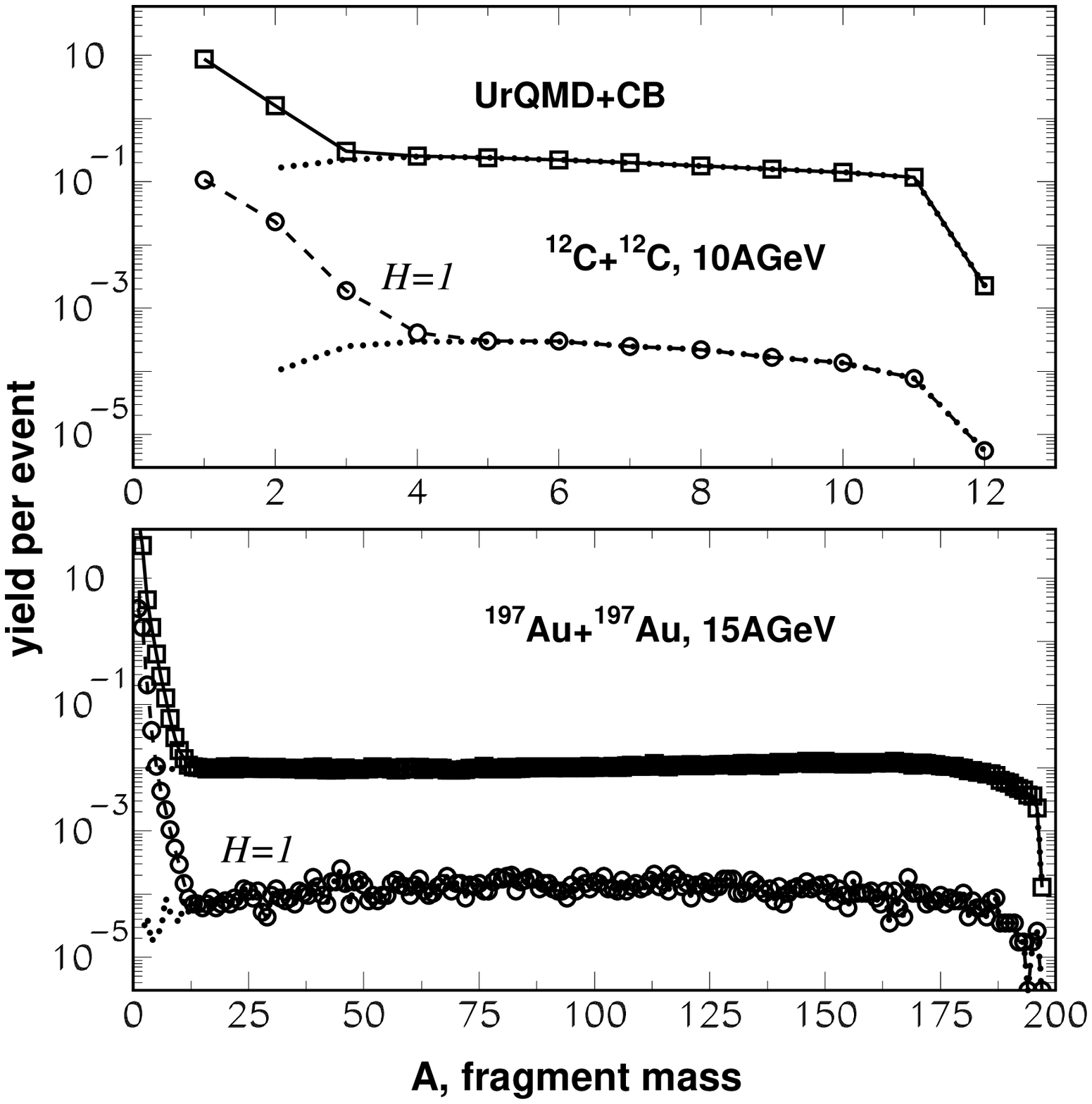}
%\includegraphics{f1-hy-wb-fig5.eps}
%\epsfig{figure=fig1.eps,width=6.8cm}
%\epsfig{f1-hy-wb-fig5.eps,width=6.8cm}
%\end{center}
\end{minipage}
\hfill
\begin{minipage}[t]{65mm}
\vspace{-73mm}
%\hspace{65mm}
\hspace{5mm}
\includegraphics[width=6.6cm]{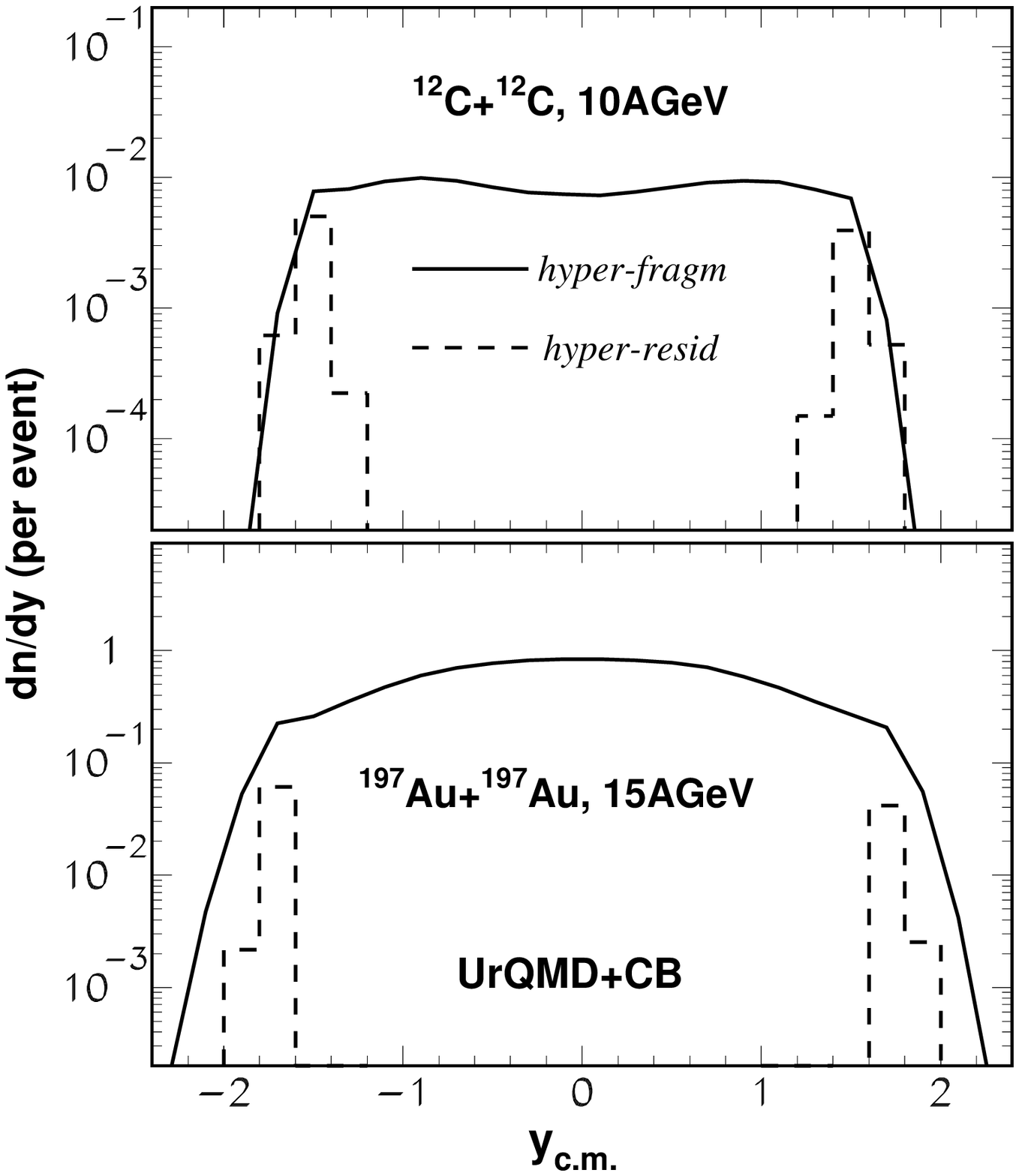}
\end{minipage}
  \caption{
The UrQMD plus CB model calculations of Au + Au and C + C ion 
collisions integrated over all impact parameters. 
%(see Ref.~\cite{Bot15}). 
The laboratory energies, projectile and targets are noted in panels. 
Left panels: Yields (per one inelastic event) of normal fragments 
(solid lines with squares) and hyper-fragments with one captured $\Lambda$ 
(notation H=1, dashed lines with circles) versus their mass number (A). 
The dotted lines present the corresponding fragments originated from the 
projectile/target spectator residues. 
Right panels: Rapidity distributions (in the center of mass system, 
$y_{c.m.}$) of all produced hyper-fragments (solid lines) and, separately, 
the hyper-residues (dashed lines) \cite{Bot15}. 
}
\end{figure}
% end my figure 

In Fig.~1 we show our theoretical predictions on the fragment 
and $\Lambda$-hyper-fragment 
production for the gold--gold and carbon--carbon relativistic collisions. 
The dynamical reaction stage is calculated with the Ultrarelativistic Quantum 
Molecular Dynamics (UrQMD) model, which is followed by the coalescence of 
baryons (CB) model \cite{Bot15}. The suggested mechanisms are the following: 
Nucleons of the projectile and target interact with binary collisions leading 
to the production of new particle including resonances. As a result of primary 
and secondary interactions between the particles and other nucleons many 
hyperons can be created. These hyperons can be captured by spectator 
residues consisting of non-interacting (spectator) nucleons, therefore, big 
pieces of hypermatter around normal nuclear density are produced. Hyperons 
can also be captured by other free baryons which are stochastically 
occurred in their vicinity at the dynamical stage. Such a condensation-like 
process is responsible for light clusters. In these cases we expect mostly 
the production of excited fragments, since the baryon capture in the 
ground states has a very low probability according to the reaction phase 
space. As was established in studies 
of normal nuclear matter the excitation energy of big spectator residues 
may reach several MeV per nucleon \cite{Bot17,Ogu11}. 
We obtain very broad mass distributions of both normal fragments and 
hyper-fragments, see Fig.~1. There are a lot of large fragments corresponding 
to the spectator residues. They concentrate in the regions of the 
projectile/target rapidities. The coalescence is 
mainly responsible for the lightest fragments and gives a contribution 
at all rapidities. The yields of coalescence fragments fall 
exponentially with $A$.  We believe 
light hyper-fragments can be used for studying hyperon and nucleon 
interactions. Whereas moderately excited big hyper-fragments are 
suitable for the investigation of hyper-matter properties. 
Important features of hypernuclei production in relativistic ion collisions 
related to the projectile/target hyper-residues were discussed in 
Refs.~\cite{Bot07,giessen,Bot11,Bot13,Buy13,Bot15,Bot17,Buy18}.

\section{De-excitation of coalescence fragments}

\vspace{-8mm}

\begin{figure}
\includegraphics[width=12.5cm]{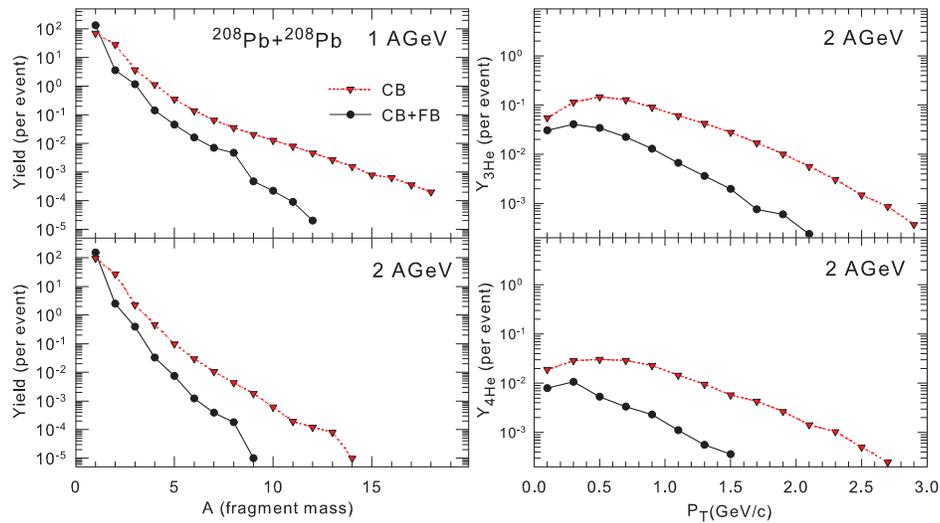}
%\vspace{2.5cm}
\caption{Fragments formed from participant nucleons in the midrapidity 
region in lead on lead collisions at projectile energies of 1 and 2 GeV 
per nucleon. The dynamical stage and participant nucleons were calculated 
with DCM. Left panels: mass yields of excited fragments as obtained with 
CB model, and ones after de-excitation of these 
fragments via Fermi-break-up (FB) process. Right panels: transverse 
momenta distributions of $^3$He (top) and $^4$He (bottom) fragments, 
before and after the de-excitation of fragments.}
\end{figure}

In this proceedings we concern the de-excitation process of light 
coalescence fragments produced abundantly in the midrapidity 
zone. Namely this zone is under study with the modern detectors 
\cite{star,alice}. We show in Fig.~2 that at intermediate projectile energies 
around 1 GeV per nucleon primary coalescence clusters including tens of 
nucleons can be 
formed, and they undergo the following de-excitation. Here all dynamical 
participant baryons were generated within the Dubna Cascade Model (DCM) 
\cite{Bot17}, while for the secondary de-excitation of clusters the 
statistical Fermi-break-up (FB) model was adopted, which was generalized 
for hypernuclei \cite{Bot13}. It is expected that the fragments 
become smaller after de-excitation (left panels). In addition, their 
isospin content changes and this can be seen in particle correlations. 
Important unstable states can be investigated in this way. 
We have also 
found an interesting isotope effect: In central collisions of big nuclei at 
low energies (less than 400 MeV per nucleon) the yield of $^4$He nuclei 
can be larger than $^3$He ones as a result of this de-excitation. (Note, that 
within a pure coalescence picture the yields of small clusters are always 
larger than big ones.) However, if we 
increase the projectile energy the yields of $^3$He become again larger 
than $^4$He ones, since the primary hot fragments will have smaller sizes. 
The kinetic energies of fragments 
change after de-excitation too. As seen from the right panels for helium 
particles the momentum distributions become steeper and maximums are 
shifted toward low energies. This is a consequence of that large primary 
fragments are formed by coalescence from slow baryons. We believe the 
extension of the coalescence towards hot clusters and their de-excitation is 
the qualitatively new development and it is consistent with the reaction 
physics.

%\paragraph{Notes and Comments.}

\vspace{3mm}

A. Botvina acknowledges the support 
of BMBF (Germany). N.B. acknowledges the Turkish Scientific
and Technological Research Council of Turkey (TUBITAK)
support under Project No. 118F111. M.B. and N.B. acknowledges that the 
work has been performed in the framework of COST Action CA15213 THOR.


\end{document}